\documentclass[journal,twoside,web]{ieeecolor}
\usepackage{generic}
\usepackage{comment}
\usepackage{hyperref}
\usepackage{amsmath,amssymb,amsfonts}
\usepackage{graphicx}
\usepackage{textcomp}
\usepackage{bm}
\usepackage{siunitx}
\usepackage{caption}
\usepackage{orcidlink}
\usepackage{subcaption}
\usepackage[backend=biber,style=ieee, minbibnames=1, maxbibnames=2]{biblatex}

\usepackage[T1]{fontenc}

\def\BibTeX{{\rm B\kern-.05em{\sc i\kern-.025em b}\kern-.08em
    T\kern-.1667em\lower.7ex\hbox{E}\kern-.125emX}}
\begin{document}
\title{Avoiding Post-processing with Event-Based Detection in Biomedical Signals}
    \author{
    Nick Seeuws\, \orcidlink{0000-0002-7024-3417},
    Maarten De Vos\, \orcidlink{0000-0002-3482-5145} , and 
    Alexander Bertrand\, \orcidlink{0000-0002-4827-8568}
\thanks{This project has received funding from the European Research Council (ERC) under the European Union’s Horizon 2020 research and innovation programme (grant agreement No 802895) and from the Flemish Government (AI Research Program).}
\thanks{N. Seeuws , A. Bertrand and M. De Vos are with the Dept. of Electrical Engineering (ESAT), Stadius Center for Dynamical  Systems, Signal Processing and Data Analytics (STADIUS), KU Leuven, Kasteelpark Arenberg 10, B-3001 Leuven, Belgium}
\thanks{M. De Vos is also with the Dept. of Development and Regeneration, Faculty of Medicine, KU Leuven}
\thanks{A. Bertrand and N. Seeuws are affiliated to Leuven.AI - KU Leuven institute for AI, B-3000, Leuven, Belgium.}
\thanks{This work has been submitted to the IEEE for possible publication. Copyright may be transferred without notice, after which this version may no longer be accessible.}
}

\maketitle

\begin{abstract}
\textit{Objective:}
Finding events of interest is a common task in biomedical signal processing. The detection of epileptic seizures and signal artefacts are two key examples.
Epoch-based classification is the typical machine learning framework to detect such signal events because of the straightforward application of classical machine learning techniques. Usually, post-processing is required to achieve good performance and enforce temporal dependencies. Designing the right post-processing scheme to convert these classification outputs into events is a tedious, and labor-intensive element of this framework.
\textit{Methods:}
We propose an event-based modeling framework that directly works with events as learning targets, stepping away from ad-hoc post-processing schemes to turn model outputs into events.
We illustrate the practical power of this framework on simulated data and real-world data, comparing it to epoch-based modeling approaches.
\textit{Results: }
We show that event-based modeling (without post-processing) performs on par with or better than epoch-based modeling with extensive post-processing.
\textit{Conclusion: }
These results show the power of treating events as direct learning targets, instead of using ad-hoc post-processing to obtain them, severely reducing design effort.
\textit{Significance}
The event-based modeling framework can easily be applied to other event detection problems in signal processing, removing the need for intensive task-specific post-processing.
\end{abstract}


\section{Introduction}
\label{section_introduction}

Machine learning has become a popular approach to solve biomedical signal processing problems, for example for epileptic seizure detection \cite{jingwei,vandecasteele2020visual,neureka,ansari2019neonatal}, the detection of sleep events \cite{li2021deepsleep,kulkarni2019deep,olesen2021msed,chambon2019dosed,yeo2021respiratory,urtnasan2020automatic,varon2015novel,erdenebayar2019deep} and sleep stages \cite{phan2019seqsleepnet}, and the detection of signal disturbances, also known as signal artefacts \cite{moeyersons2019artefact,moeyersons2021artefact,behar2013ecg,zhao2022quantitative,malafeev2019automatic,webb2021automated}. 

A specific group of tasks in biomedical signal processing can be said to deal with \textit{events}. Conceptually, these tasks involve the detection of transitions to a signal event of interest and back to the background signal. The detection of epileptic seizures and signal artefacts are two such examples, where the seizures and artefacts are the events of interest. The manual event detection process involves scrolling through signal recordings (potentially representing multiple hours of signal) and logging the start and stop times of the events. Designing tools to aid human annotators requires linking raw signal recordings with a set of $(t_{start}, t_{stop})$ tuples. 

The prototypical machine learning approach to detect such events in biomedical signals involves segmenting a signal recording into distinct epochs and predicting a label for each individual epoch \cite{craik2019deep}. Deciding on the epoch duration is part of the design process, and also heavily relies on the machine learning model that is used. An extreme case is where a prediction is made on the level of individual time samples (which could be viewed as single-sample epochs), as often done in U-Net-like architectures \cite{neureka, li2021deepsleep}.

Converting predictions for every individual epoch into a $(t_{start}, t_{stop})$ tuple spanning a continuous event, possibly across multiple epochs, involves extensive post-processing \cite{vandecasteele2020visual,jingwei,neureka}. This post-processing is usually not learned from data, but designed based on expert knowledge. If epochs are fed to a machine learning model independently the post-processing stage is also responsible for encoding temporal dependencies inherent to time series processing into the final output. This responsibility makes the post-processing stage a crucial ingredient with a tedious design process.

In this work, we introduce \textit{event-based} modeling as an alternative paradigm (instead of epoch-based modeling). We will illustrate how this method bypasses the need for a tedious ad-hoc design of a proper post-processing stage. Inspired by works in visual object detection, our method encodes events of interest using the events' center and duration. Both are predicted jointly by a single deep learning model. We consider this to be an \textit{event-based} approach to biomedical signal analysis.

Event-based modeling reduces many separate design tasks to the design of a single neural network, without a need to carefully tune pre- or post-processing steps (as both are directly learned by the model). Encoding training events involves mapping the different events to an \textit{event center} and \textit{duration} signal, after which the model can be trained end-to-end without post-processing. Explicitly modeling events, combined with the end-to-end nature of deep learning, encourages the model to properly learn the full character and diversity of target events. Crucially, we can easily cope with a large variability in event duration.

To summarize, our key contributions are as follows:
\begin{itemize}
	\item Introduction of a generic \textit{event-based} modeling framework for biomedical signal processing
	\item An event-detection algorithm that does not require tailored, task-specific post-processing (in contrast to most epoch-based approaches)
	\item Due to its end-to-end nature, our algorithm learns the full character and diversity of target events of variable duration
\end{itemize}

Section \ref{section_methods} introduces our event-based framework and discusses our experiments. Section \ref{section_results} shows results of these experiments. Section \ref{section_discussion} discusses these results and explains benefits and drawbacks of using an event-based framework. Section \ref{section_conclusion} concludes the paper.

\begin{figure*}[ht!]
	\centering
	\includegraphics*[width=0.8\textwidth]{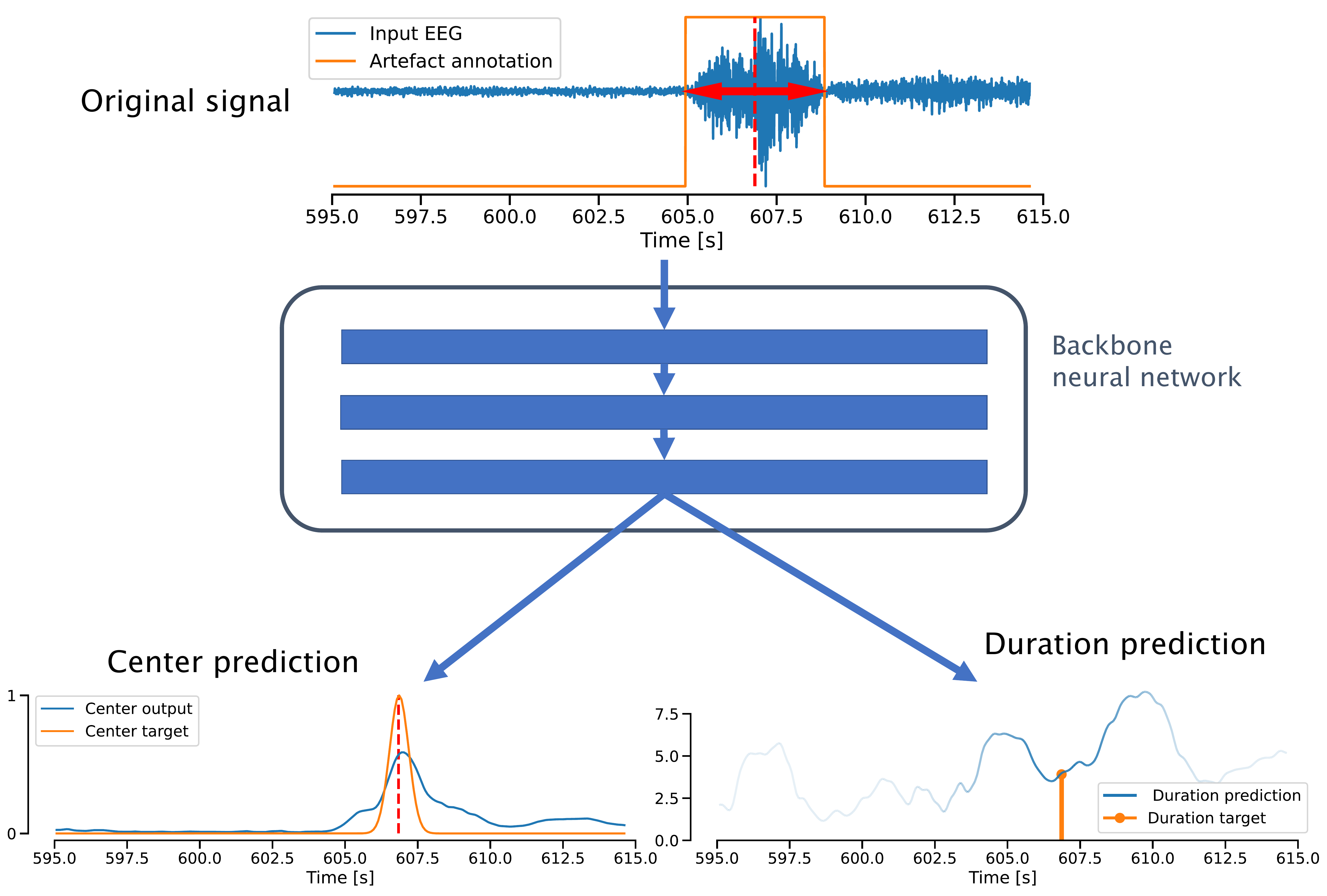}
	\caption{Event-based modeling overview. The input EEG signal at the top contains a single artefact event, annotated in orange (the event's center and duration are highlighted in red). The neural backbone can be any sequence-to-sequence model. Two independent convolutional layers output the \textit{center} and \textit{duration} signals (in blue). The training targets for both signals are plotted in orange. Training the \textit{center} signal involves comparing it to the entire target signal. The \textit{duration} signal is only trained and evaluated at event centers. Note that this example only spans \SI{20}{\second} and contains a single, easy to detect event. With the right backbone, the approach can process longer inputs and detect more events at once.}
	\label{fig:algo}
\end{figure*}
\section{Methods}
\label{section_methods}

\subsection{Event-based modeling}
In this subsection, we describe the event-based framework conceptually, and we refer to Figure \ref{fig:algo} for a schematic illustration of the different concepts introduced in this subsection.

\subsubsection{Encoding and decoding events}
Our event-based framework represents events by their center point and duration. A signal recording is used as input, and the goal is to produce center and duration predictions across the length of the recording. Both outputs (i.e., center and duration) are treated as "signals", in the sense that they span the entire input. This situates our approach in the area of "sequence to sequence" learning, similar to the works of \cite{phan2019seqsleepnet, neureka, li2021deepsleep}. The \textit{center signal} indicates whether a point in time corresponds to the center of an event. The \textit{duration signal} is used to represent an event's duration \textit{if that point in time would be a center point}. Note that this duration signal is meaningless at time points far away from a center point. Our approach is inspired by CenterNet \cite{zhou2019objects} for object detection in images. This image object detection model predicts centers of detected objects, and predicts object sizes at those specific centers.

At inference time, the center and duration outputs are decoded by searching for the peaks in the center signal. The detected event centers are then represented by the different peaks, and the confidence level for each event is displayed by the specific signal values at the corresponding peaks. \footnote{This confidence is expressed on a relative scale, and does not necessarily represent detection probabilities like they do for classification problems. These signal values should be interpreted as "confidence scores", e.g., a model is more confident in its prediction for an event with score 0.8 than it is for an event with 0.3 as score.} For each detected center, the predicted event duration can then be found in the duration signal at that specific point.

\subsubsection{Losses}
\paragraph{Training targets}
To train the center and duration predictions, training targets need to be defined. Center prediction is treated as a sample-based classification task, similar to the approaches of \cite{neureka, li2021deepsleep}. In contrast to most sample-based classification approaches, some slack should be allowed on the center targets. Predicting a center that is just a few samples off-target is better than, e.g., predicting a false event in an hour of background signal, and thus should be penalized less. The weighting method of \cite{zhou2019objects} for object detection is modified and applied to event center prediction. For an event with ground truth center $t^*$, the target center signal is defined as 
\begin{equation}
	c(t) = \exp ( - \frac{(t - t^*)^2}{2 \sigma^2} )
	\label{eq:center}
\end{equation}
with $\sigma$ depending on the target event's duration. Following \cite{liu2020training}, the hyperparameters are set as $\sigma = \alpha d / 6$, with $\alpha = 0.5$ and $d$ the event's actual duration, measured in terms of \textit{time points}. These hyperparameters can be adjusted for a specific use case to specify how precise a model should be in its center predictions during training (e.g, a larger value for $\sigma$ allows for more "slack" on center predictions.) In this work, we use the values of \cite{liu2020training} as-is, and leave the tuning of these hyperparameters as future work. In the case of multiple events in a signal, these center target signals as in Eq. \ref{eq:center} are defined for each event independently, and are combined by taking the maximum target value at every time point.

The duration targets only need to be defined at the target centers $t^*$, since only duration predictions at ground-truth center points $t^*$ will be considered in the duration loss. As presented, our approach works with a maximum duration it can predict. This can either be set to the maximum duration in a given data set, or determined with expert information (e.g.,\textit{"What can reasonably be expected as an upper bound for these events?"}). The duration predictions, similar to the center predictions, are constrained to the range $[0, 1]$ (the target durations are divided by the predefined maximum duration.) For every target event in a data set, the duration signal value at the event's center point is set to the event's normalized duration.

\paragraph{Training losses}
Center prediction is trained using focal loss \cite{lin2017focal}, in the modified form of \cite{law2018cornernet}. The full center signal prediction loss $L_c$ for an input signal containing $N$ events, center prediction $c'(t)$, and target center signal $c(t)$ is defined as:
\begin{equation}
	L_c = - \frac{1}{N} \sum_t 
	\begin{cases} (1 - \alpha_c) (1 - c')^\alpha \log (c') & \text{if} \: c = 1 \\ 
	a_c (1 - c)^\beta c'^\alpha \log (1 - c') & \text{otherwise}
	\end{cases}
\end{equation}
(with the dependence on $t$ of $c'(t)$ and $c(t)$ dropped for legibility). Hyperparameters are set as $\alpha_c = 0.1, \alpha = 2$ and $\beta = 4$ following \cite{lin2017focal,law2018cornernet}. The original focal loss is a classification loss where mistakes are adaptively penalized based on the model's confidence using the factors exponentiated with $\alpha$.  With the modification of \cite{law2018cornernet}, false alarms close in time to the target center $t^*$ are penalized less than false alarms further away (using the factor $(1 - c) ^\beta)$ and the exponential in Eq. \ref{eq:center}). \footnote{This form of the focal loss can cause numerical instabilities when performing gradient descent during training. To avoid instabilities, one needs to rewrite the loss in terms of the logits instead of sigmoid inputs. Refer to Appendix \ref{app:stable} for more details.}

Duration predictions are trained using \textit{Intersection over Union} (IoU) as loss. IoU is a popular loss formulation in object detection \cite{liu2020training}. Crucially, IoU is based on \textit{relative} duration errors, ensuring that the batch loss will not be dominated by long events. Calculating the intersection and union of predicted and target events can be simplified due to the use of the target event's center. The duration loss $L_d$, for $N$ events, set of known target points $\mathcal{T}$, predicted duration signal $d'(t)$ and target $d(t)$, is formulated as 
\begin{equation}
	L_d = \frac{1}{N} \sum_{t^* \in \mathcal{T}}
	\frac{\min [d'(t^*), d(t^*)]}
	{\max [d'(t^*), d(t^*)]}
\end{equation}
Note that the predicted duration signal $d'(t)$ is only evaluated at the center points, i.e., the value of $d(t)$ and $d'(t)$ has no meaning at points which are not treated as center points, even though the network will produce an output $d'(t)$ for every time point $t$.

The center prediction loss $L_c$ and duration loss $L_d$ are combined into the full loss $L$ as a weighted sum, $$L = L_c + \lambda_d L_d$$, where $\lambda_d$ is a hyperparameter to control the relative influence of the two tasks (center and duration prediction). In our experiments, both loss terms become approximately equal in magnitude by setting $\lambda_d = 5$. This value can be raised or lowered to increase or decrease, respectively, the influence of the duration prediction task.

\subsubsection{Backbone model}
Taking an event-based approach to modeling events does not rely on a specific backbone architecture. As discussed above, one can view it as a specific instance of sequence-to-sequence modeling. Hence, any neural network architecture that maps an input signal to an output signal can be applied in an event-based context (one would need to account for the center and duration signals by converting the architecture to produce two outputs).

In our experiments, we use U-Net-like backbones (tailored to a specific data set), ensuring we use use the same backbone architecture for the event-based and epoch-based approaches for fair comparison. This type of architecture won two recent machine learning competitions in the context of event detection in biomedical signals\cite{neureka,li2021deepsleep}. The architecture is capable of mapping an input signal (uni- or multivariate) to the desired center and duration output signals, and manages to combine global and local information of the input. Due to differences in scale of the target events in our experiments, discussed below, specific implementations of the backbones are tailored to the specific data sets. Details can be found in Appendix \ref{app:details}.

\subsection{Experiments}

\subsubsection{Measuring performance}
Measuring performance of an algorithm in the context of biomedical events is not straightforward. In the field of epileptic seizure detection, for example, multiple measures are in active use, and all focus on different aspects of "performance" \cite{ziyabari2017objective}. Broadly speaking, there are two categories of performance measures: epoch-based and event-based. Epoch-based measures treat the \textit{evaluation} of an algorithm similarly to epoch-based solutions, i.e., as separate classification for each epoch. In this case, one can rely on classical performance measures for classification problems (accuracy, precision, recall, etc.).

Event-based measures, on the other hand, take a more holistic approach to evaluation and focus on the events in question. They look at how well predictions overlap with the reference annotations, and match up predicted events with reference events. Different measures vary in how they quantify this overlap, and what they consider as "enough overlap". For example, in \cite{ziyabari2017objective} the authors discuss, among others, the \textit{any-overlap} and \textit{time-aligned event scoring} methods. Broadly speaking, the former considers a prediction to be a correct prediction if there is any temporal overlap with ground-truth events, while the latter also considers the amount of overlap between predictions and ground truth. Both measures can give different results for a single set of predictions, and it is up to the user to decide what measure best corresponds to the problem at hand. In object detection for computer vision, the community uses a generic, tunable approach to measure performance, also based on the amount of overlap between predictions and ground truth \cite{Redmon_2016_CVPR,zhou2019objects,liu2020training,lin2017focal,law2018cornernet,lin2014microsoft}. Matches between the set of predicted objects and ground-truth objects are made based on maximal overlap, measured using \textit{Intersection-over-Union} (IoU). For the prediction-ground-truth pairs obtained like this, if the overlap is higher than a specific IoU threshold, the pair counts as a correct detection. Remaining predicted and ground-truth objects are counted as false positives and false negatives respectively.

In the context of this work we elect to measure performance using an event-based measure, since designing an \textit{event-based} model is mainly relevant if the final evaluation will also take an event-based point of view. Therefore, we will use the IoU as the main performance metric throughout this paper. Additionally, deciding on true/false positives and negatives based on a tunable overlap threshold allows to elegantly make the evaluation more or less strict, depending on the problem at hand. As an illustration of the power of event-based modeling, this flexibility is desirable in the context of this paper. For example, the \textit{any-overlap} scoring of \cite{ziyabari2017objective} can be seen as a limit case of the IoU threshold going to zero, and will be applicable in use cases where the \textit{any-overlap} scoring is also relevant. On the other hand, setting a high IoU threshold is suited to evaluate algorithms when there is a high standard on the overlap between predicted and ground-truth events.

\subsubsection{Simulated events}
As a first test we simulate a data set by mixing realistic noise events in electrocardiography (ECG) signals. This ensures unambiguous annotations, which can be difficult to obtain in real-life biomedical data (e.g., in epileptic seizure detection, where the precise starting point of seizure is difficult to define). As background signal, we use the Computing in Cardiology 2017 Challenge data set \cite{clifford2017af}. This data set contains lead II ECG recordings of sinus rhythm ECG and atrial fibrillation. We randomly generate electrode artefact events with varying durations sourced from the Physionet MIT-BIH Noise Stress Test Database \cite{moody1984noise}. The artefact events are added to the background signal with varying SNR levels. To "smoothen" the transition between background and artefact, the artefacts are elementwise multiplied with a Tukey window of the same size as the artefact event. To make the task more difficult, we add short bursts of artefact signal throughout the data set (which should be ignored by the models). Full data generation details are discussed in Appendix \ref{app:generation}

We compare our event-based approach to a generic epoch-based approach. Using a U-Net-like backbone for event-based modeling allows for direct comparison with epoch-based modeling by training an actual U-Net (with a single output) in an epoch-based manner, where it is trained to produce predictions at $1/16$ the original sampling rate (making it so that the two settings use the same backbone architecture and the addition of the center and duration output is the only change). Network details are explained in Appendix \ref{app:sim_details}. The event-based model is used as-is, while the epoch-based model is used in three different settings: 
\begin{itemize}
	\item \textit{No post-processing}: To establish a baseline, we report performance of the epoch-based approach \textit{without} post-processing to gauge the impact of post-processing schemes.
	\item \textit{Median filtering}: To compensate for potential (short) false positives, which we know to be a risk due to the short "distraction" events added into the data generation process, we use median filtering after thresholding the base-model output as a first post-processing scheme. The filter length is equivalent to \SI{1}{\second}, the shortest possible duration of the target events.
	\item \textit{Morphological operations}: As a more advanced post-processing scheme, we take inspiration from morphological operations popular in the field of computer vision \cite{haralick1987image}. After training the base epoch-based model, we observed more room for improving results than median filtering. The U-Net would predict "holes" in target events which are not always fixed with median filtering (especially if they occur close to event edges), in addition to false positives. To further push post-processing, we use \textit{binary closing} (which closes holes in the foreground, i.e., a predicted event), followed by \textit{binary opening} (which removes short events, expected to be false positives) with a binary structuring element of size equivalent to \SI{1}{\second}, applied after thresholding the base-model output.
\end{itemize}

To measure performance, we use two different IoU thresholds, 0.25 and 0.75, to determine "hits" and "misses" of the two approaches. These two thresholds represent two different evaluation settings, one where the overlap between predictions and ground truth is not that important (0.25 IoU) similar to the \textit{any-overlap}, and a setting where the precise overlap is more important (0.75 IoU). We randomly generate 25 data sets to control for variability. We report the F1-score corresponding to the operating point that gives the highest score. Operating points for the event-based approach are set by including events with a confidence score higher than a certain threshold. Operating points for the epoch-based approaches are set by thresholding the model output by a certain value. We opt for the F1-score instead of looking at the full precision-recall curves since the epoch-based approaches are observed to only have a singular Pareto-optimal operating point (illustrated in Appendix \ref{app:precision_recall}). The epoch-based approaches would thus realistically only be used at this singular operating point, making it unfair to evaluate them over a full range of points. The flexibility of choosing an operating point is an additional benefit of using our event-based modeling approach, but will not be evaluated in this experiment.

\begin{figure}[t]
	\centering
	\includegraphics[width=\linewidth]{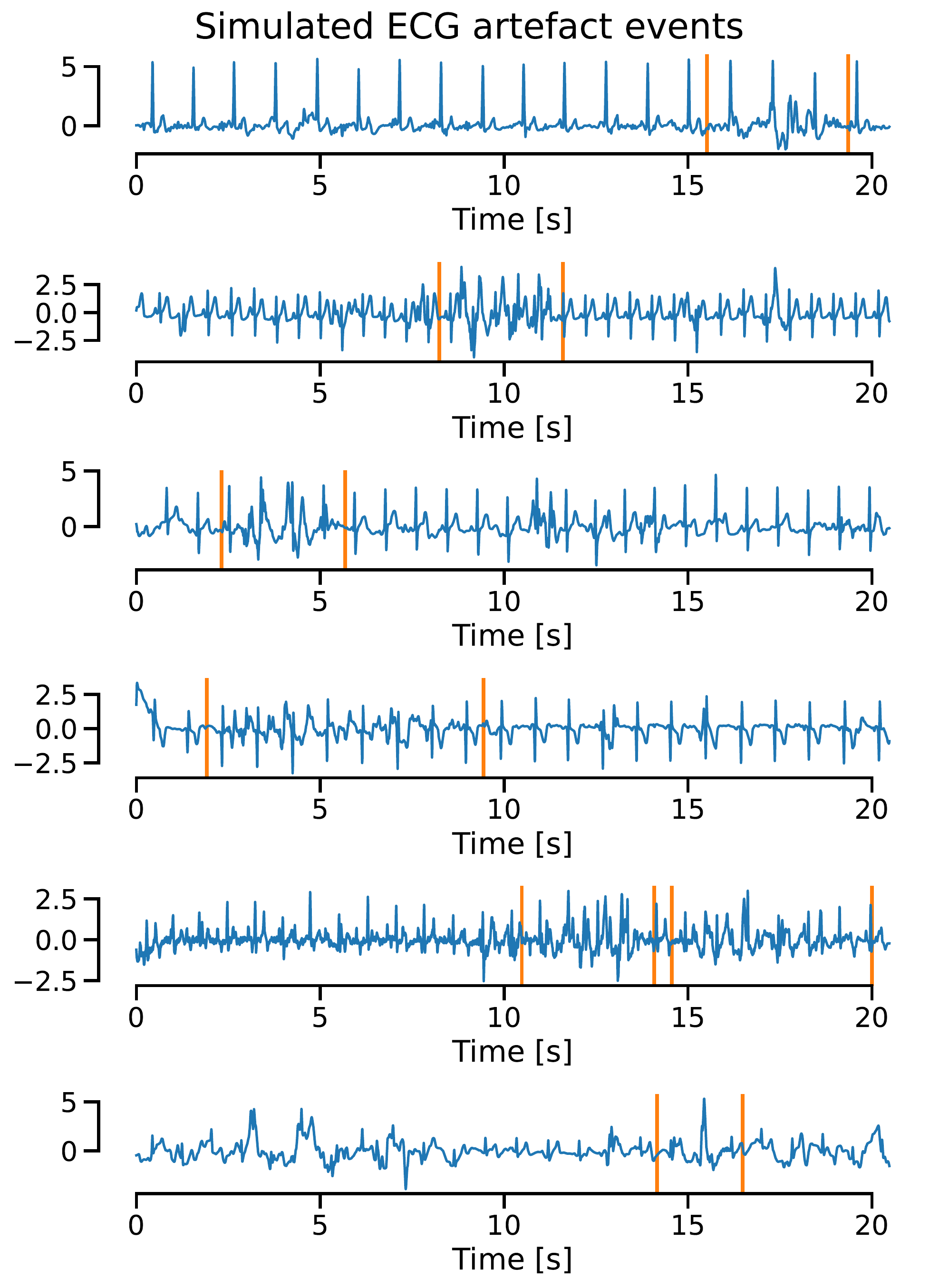}
	\caption{Examples of simulated events. The orange lines indicate the start and stop times of each target event. Signal amplitude is unitless.}
	\label{fig:simulation}
\end{figure}

\subsubsection{Real-world data}
Additionally, we show results on two real-world data sets, one containing EEG artefacts and the other one containing EEG with epileptic seizures. 

For the real-world data sets, we use 0.5 IoU as a threshold as a meet-in-the-middle metric between an any-overlap scoring and a time-aligned event scoring \cite{ziyabari2017objective}. Instead of reporting an aggregate measure like average precision, we compute precision at operating points corresponding to specific recall values (since, in a real-world setting, users would also need to decide on an operating point). Next to the precision using 0.5 IoU, we also compute the proportion of predictions that still have positive overlap with a corresponding ground-truth event, but less than 0.5 IoU. This allows for a deeper understanding of what sort of predictions the two approaches produce.

Evaluating performance solely based on IoU combines both "branches" of our event-based approach (center and duration predictions). In addition, we also evaluate the performance of both these aspects independently. To do so, predicted events that have positive overlap with their corresponding ground-truth events have their center point offsets and duration differences evaluated. In the case of the epoch-based approaches, the center point and duration are extracted directly from the events that are outputted by the post-processing stage, whereas for the event-based approach the center point and duration are directly obtained from the outputs of the neural network. We report \textit{relative} errors for center point offsets and duration errors.

\begin{figure}[t]
	\centering
	\includegraphics[width=\linewidth]{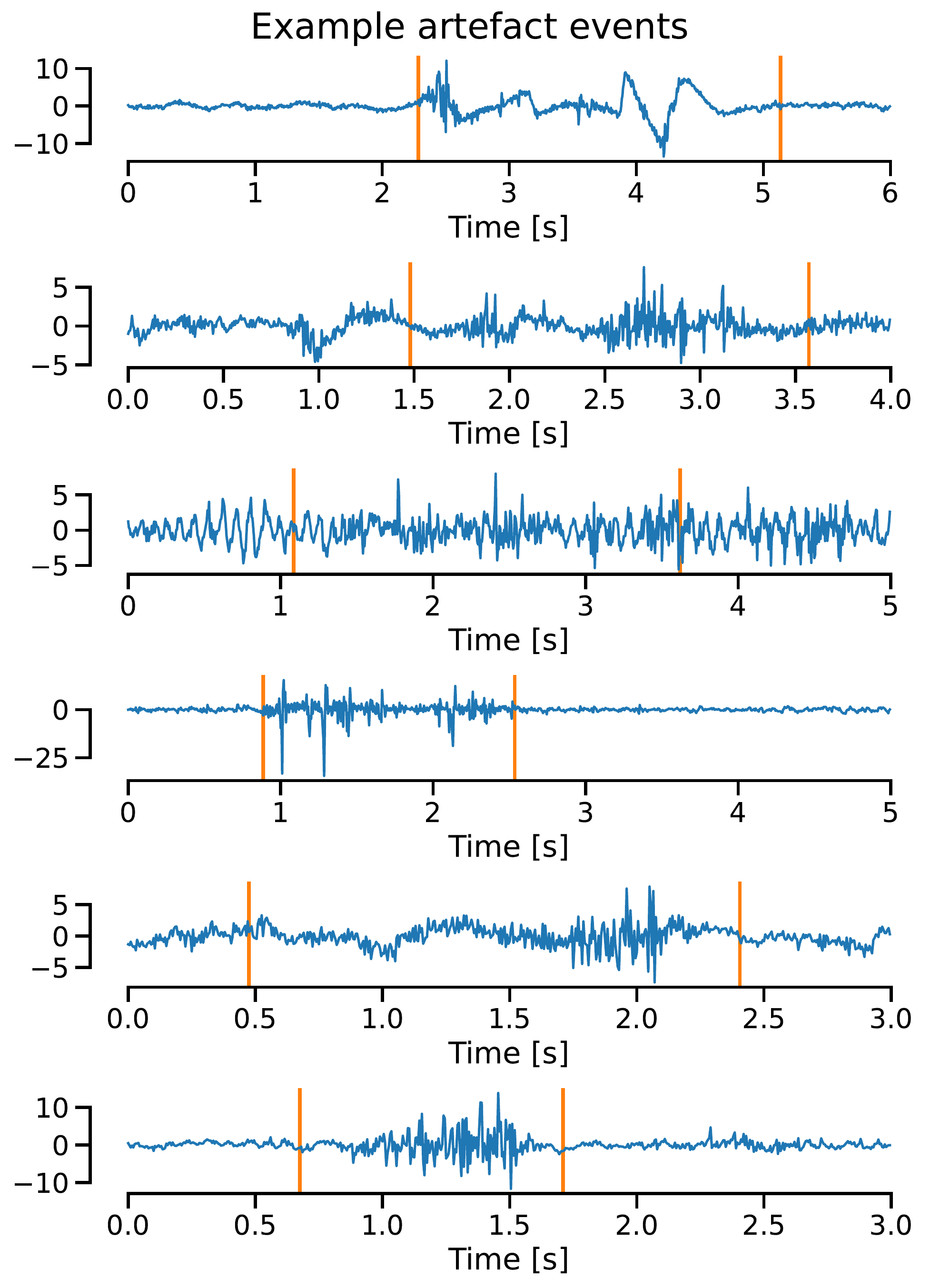}
	\caption{Examples of EEG artefact events. The orange lines indicate the start and stop times of each event, as annotated in the Temple University artefact data set. Signal amplitude is unitless.}
	\label{fig:artefact}
\end{figure}

\paragraph{EEG artefacts}
The Temple University Artefact Corpus consists of various EEG artefact events, described in \cite{hamid2020temple}. In the full data set, many multi-channel EEG recordings with channel-level annotations of artefact events are present. In this paper we focus on the muscle and chewing artefacts due to their large variability in duration. Both types are joined into a single artefact class. We train models on the individual channels of this data set to predict the channel-level artefact annotations. The recordings are divided into training, validation, and test sets, making sure that recordings of the same individual are not split among the sets. Example artefact events can be found in Figure \ref{fig:artefact}.

The event-based and epoch-based model both use the same backbone architecture. Post-processing for the epoch-based model is done with a median filter with filter length of \SI{0.1}{\second}. Network details can be found in Appendix \ref{app:details}.

\begin{figure}[t]
	\centering 
	\includegraphics[width=\linewidth]{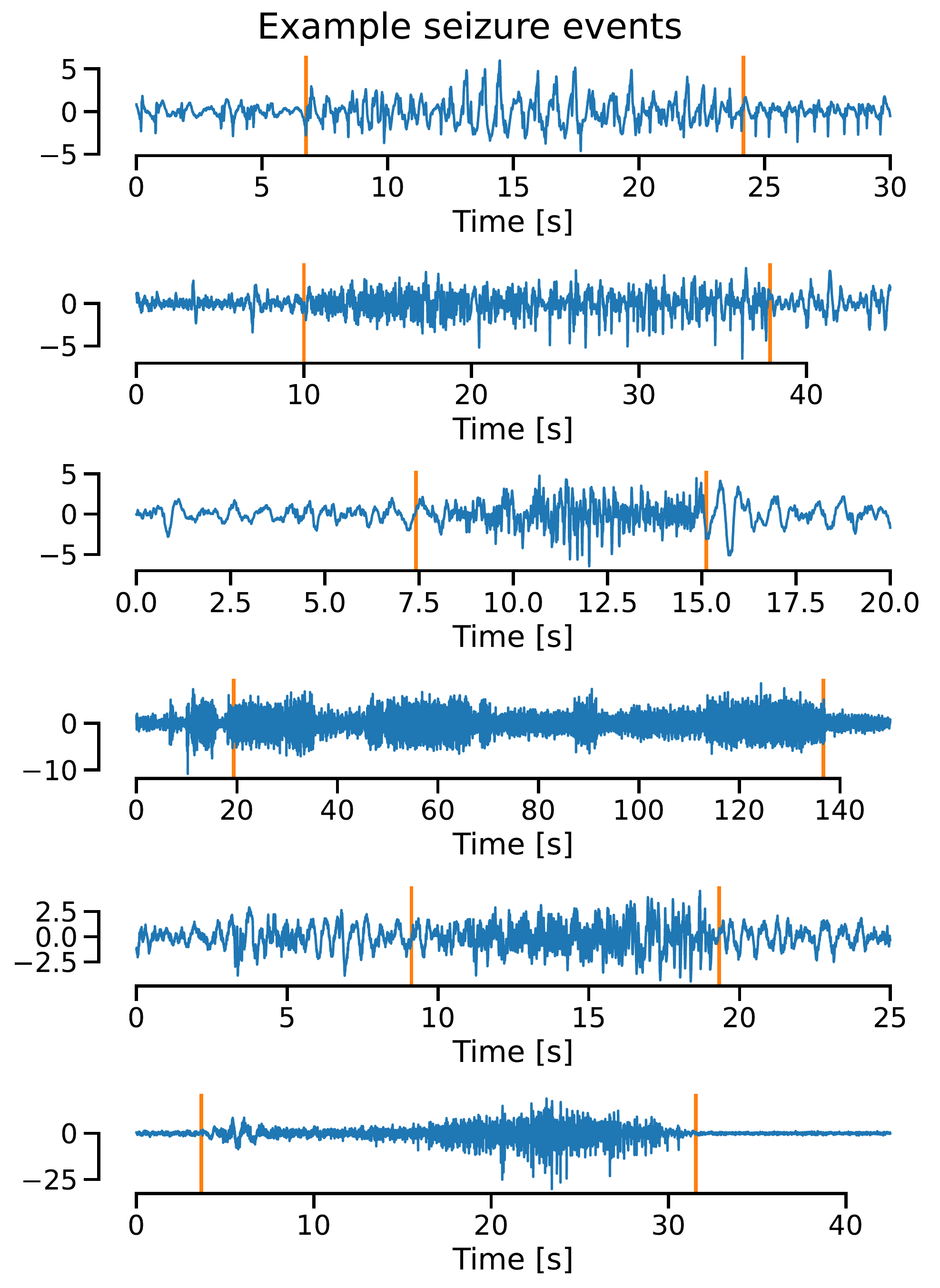}
	\caption{Example of seizure events. These examples are single channels taken from multichannel EEG. Orange lines indicate the start and stop times of each event, as annotated in the Temple University seizure data set. Signal amplitude is unitless.}
	\label{fig:seizure}
\end{figure}

\paragraph{Epileptic seizures}
The second real-world data set is the Temple University Seizure Corpus containing epileptic seizures \cite{shah2018temple}. The data set is made up of multi-channel EEG recordings, with epileptic seizures annotated at a general level (only indicating at which point in time a seizure occurs, not on which channel(s)). Example seizure events can be found in Figure \ref{fig:seizure}

As comparison, we use the approach of \cite{neureka}, which won an international seizure detection challenge on this specific data set \cite{neureka_challenge}. The approach consists of an epoch-based learning task with a U-Net architecture combined with extensive post-processing, tailored to seizure detection and the data set in question. For the sake of a fair comparison, we use the same backbone architecture as \cite{neureka} in our event-based model, yet without the post-processing stage, showcasing the performance of our generic approach compared to an epoch-based algorithm with heavily tailored post-processing. Network details can be found in Appendix \ref{app:details}.

\section{Results}
\label{section_results}

\subsection{Simulated events}

\begin{figure*}[t]
	\centering
	\begin{subfigure}[t]{0.49\textwidth}
		\centering
		\includegraphics[width=\textwidth]{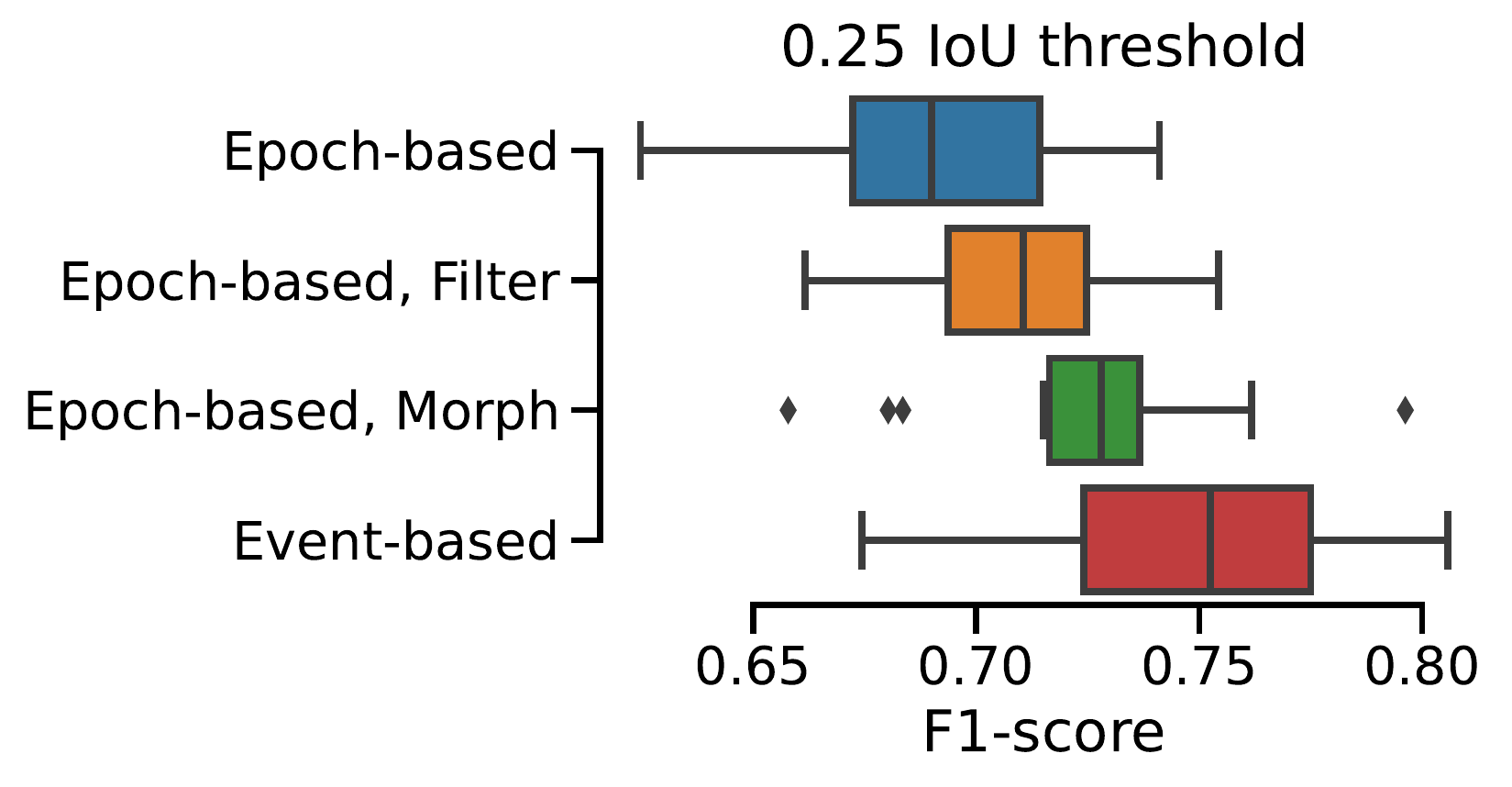}
		\caption{}
		\label{fig:simulated_025}
	\end{subfigure}
	\hfill
	\begin{subfigure}[t]{0.49\textwidth}
		\centering
		\includegraphics[width=\textwidth]{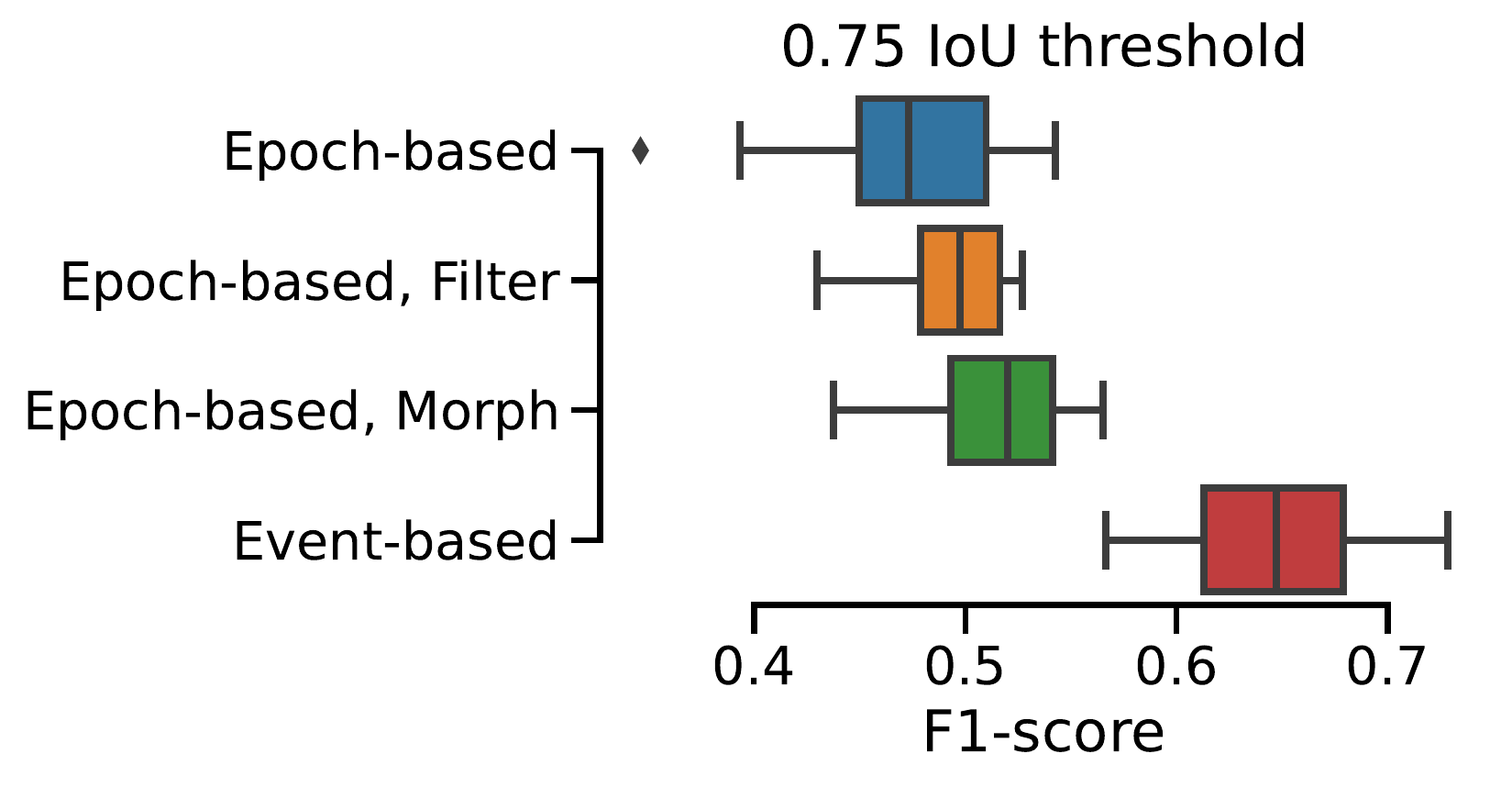}
		\caption{}
		\label{fig:simulated_075}
	\end{subfigure}
	\caption{Summary results of different training runs on simulated ECG artefacts. We evaluate performance using 0.25 and 0.75 IoU as the "threshold" to compute true/false positives and negatives to reflect a lenient and more strict evaluation setting. "Epoch-based" is an epoch-based approach \textit{without} post-processing, "Epoch-based, Filter" is an epoch-based approach with median filtering, "Epoch-based, Morph" is an epoch-based approach combined with morphological operations, and "Event-based" is our event-based approach.}
	\label{fig:simulated}
\end{figure*}

Results for the simulated events are shown in Figure \ref{fig:simulated}. For 0.25 IoU (a more lenient setting), the event-based approach outperforms all epoch-based approaches. Additionally, one can see the impact of post-processing in an epoch-based setting. Using a median filter improves upon an approach without post-processing. Post-processing based on morphological operations, a more intensive and task-specific scheme, further improves results. 

For 0.75 IoU (a more strict setting), a similar pattern can be observed. The event-based approach outperforms the epoch-based approaches, and post-processing improves results for the epoch-based approaches. However, the effect of post-processing is smaller.

\subsection{Real-world data}

\begin{figure*}[t]
	\centering
	\begin{subfigure}[t]{\textwidth}
		\centering
		\includegraphics[width=\textwidth]{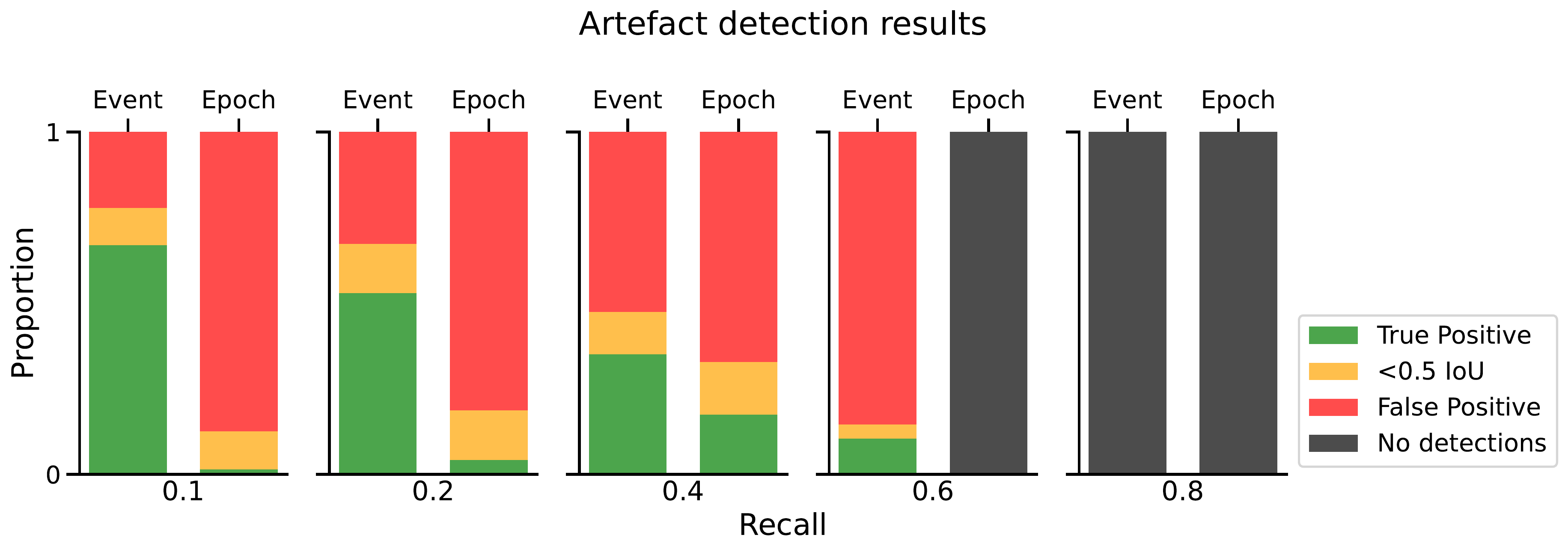}
		\caption{}
		\label{fig:tuar_sampled}
	\end{subfigure}
	\hfill
	\begin{subfigure}[t]{\textwidth}
		\centering
		\includegraphics[width=\textwidth]{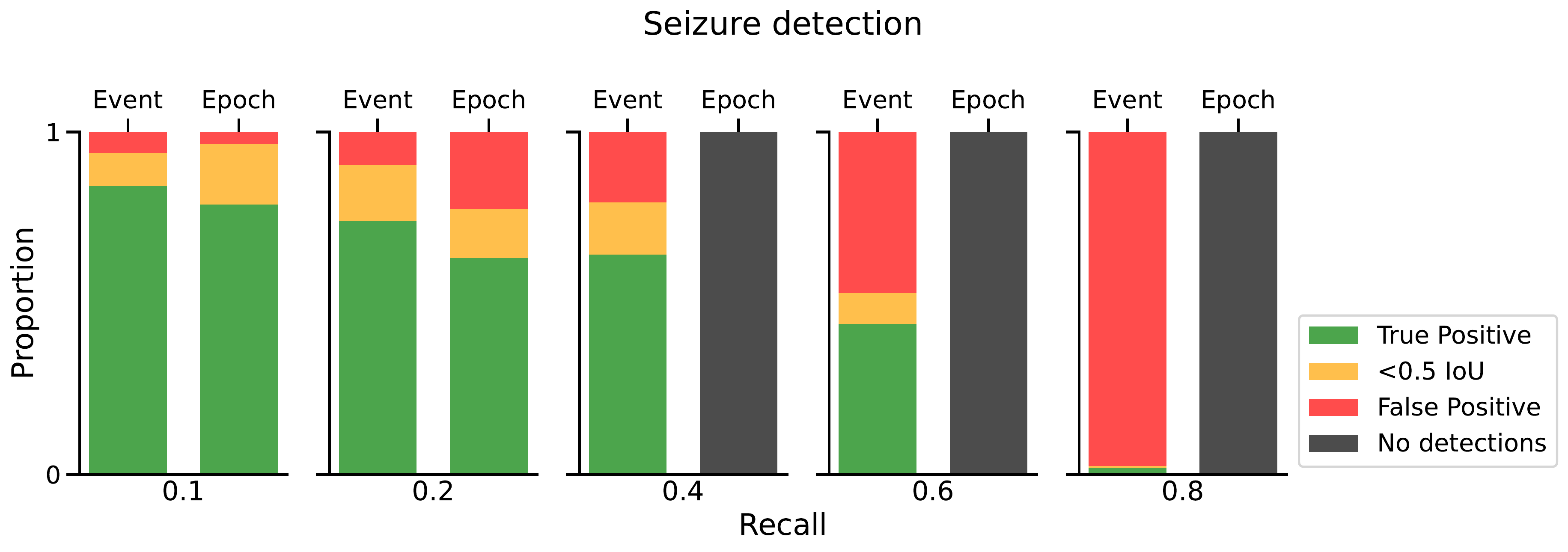}
		\caption{}
		\label{fig:tusz_sampled}
	\end{subfigure}
	\caption{Detection results for real-world EEG artefact and seizure events at different recall levels. \textit{True positive} detections correspond to $\text{IoU} \geq 0.5$. The proportion of predictions that have $\text{IoU} < 0.5$ but still have positive overlap are indicated in orange. \textit{False positive} detections have no overlap with ground-truth events. \textit{No detections} indicates that the algorithm cannot detect events at that recall level.}
	\label{fig:precrec}
\end{figure*}

Artefact and seizure detection precision is displayed in Figure \ref{fig:precrec} for different recall levels (computed using the IoU-based detection criterion, with 0.5 as threshold). Overlaps over 0.5 IoU are counted as \textit{true positive} detections. Detected events having less than 0.5 IoU with a matched reference event, but which still have positive overlap, are indicated as <0.5 IoU. How well the approaches predict the duration of a detected event is gauged by the comparison between the true positive and <0.5 IoU detections. Predicted events that do not correspond to a ground-truth event are counted as \textit{false positives}. A network that cannot detect events at a specific recall level is indicated as \textit{No detection}.

For most recall levels, the event-based approach is outperforming the epoch-based one. Additionally, more events are found using the event-based approach compared to the epoch-based case, shown by the former's higher recall level. Note that both approaches do not reach 100 \% recall. Unlike for binary classification, not all targets (events) get detected by moving the decision threshold to zero because of the IoU threshold to count predicted events as true positives.

The seemingly rising precision-recall curve for the epoch-based approach in the artefact data set is an unintuitive finding. Normal precision-recall curves are expected to show high precision at low recall, and low precision at high recall. This behavior is not shown in this case. The behavior can be explained by the decoding process of the epoch-based predictions. To predict events at specific recall levels, \textit{confidence thresholds} need to be varied, which are set to correspond to a specific cutoff of the output to distinguish between \textit{event} and \textit{background}. Because of some noise in this output signal, two events can easily be detected as a single long event if the threshold is low. On the other side, a single event can easily be split into multiple shorter ones if the threshold is rather high. This sensitivity to a cutoff, and the fact that a 100 \% recall cannot be achieved, can result in an irregular precision-recall curve with a single Pareto-optimal operating point. Similar behavior is observed for the simulated events, as illustrated in Appendix \ref{app:precision_recall}.

\begin{figure*}[t]
	\centering
	\begin{subfigure}[t]{0.4\textwidth}
		\centering
		\includegraphics[width=\textwidth]{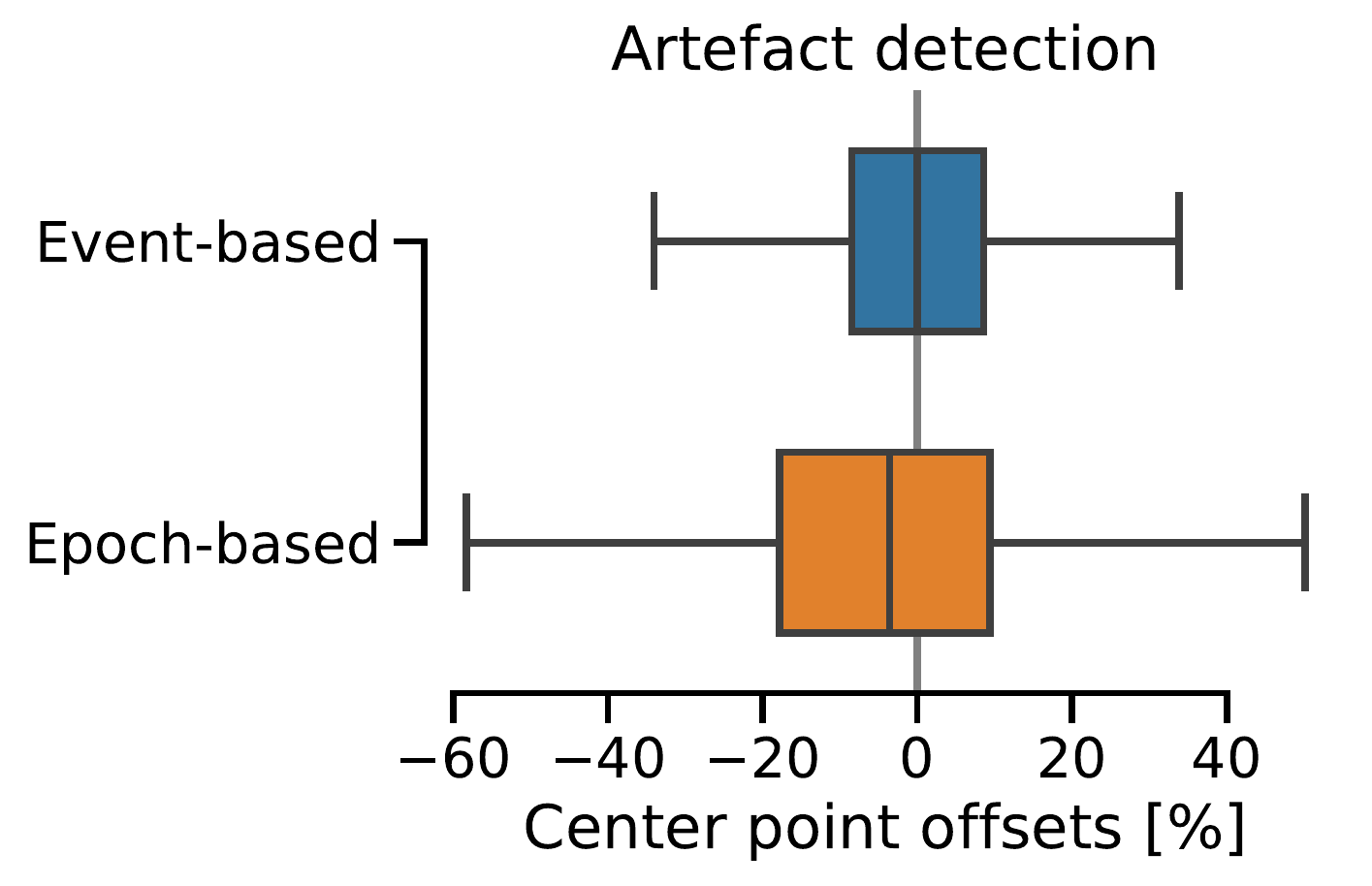}
		\caption{}
		\label{fig:tuar_center}
	\end{subfigure}
	\hfill
	\begin{subfigure}[t]{0.4\textwidth}
		\centering
		\includegraphics[width=\textwidth]{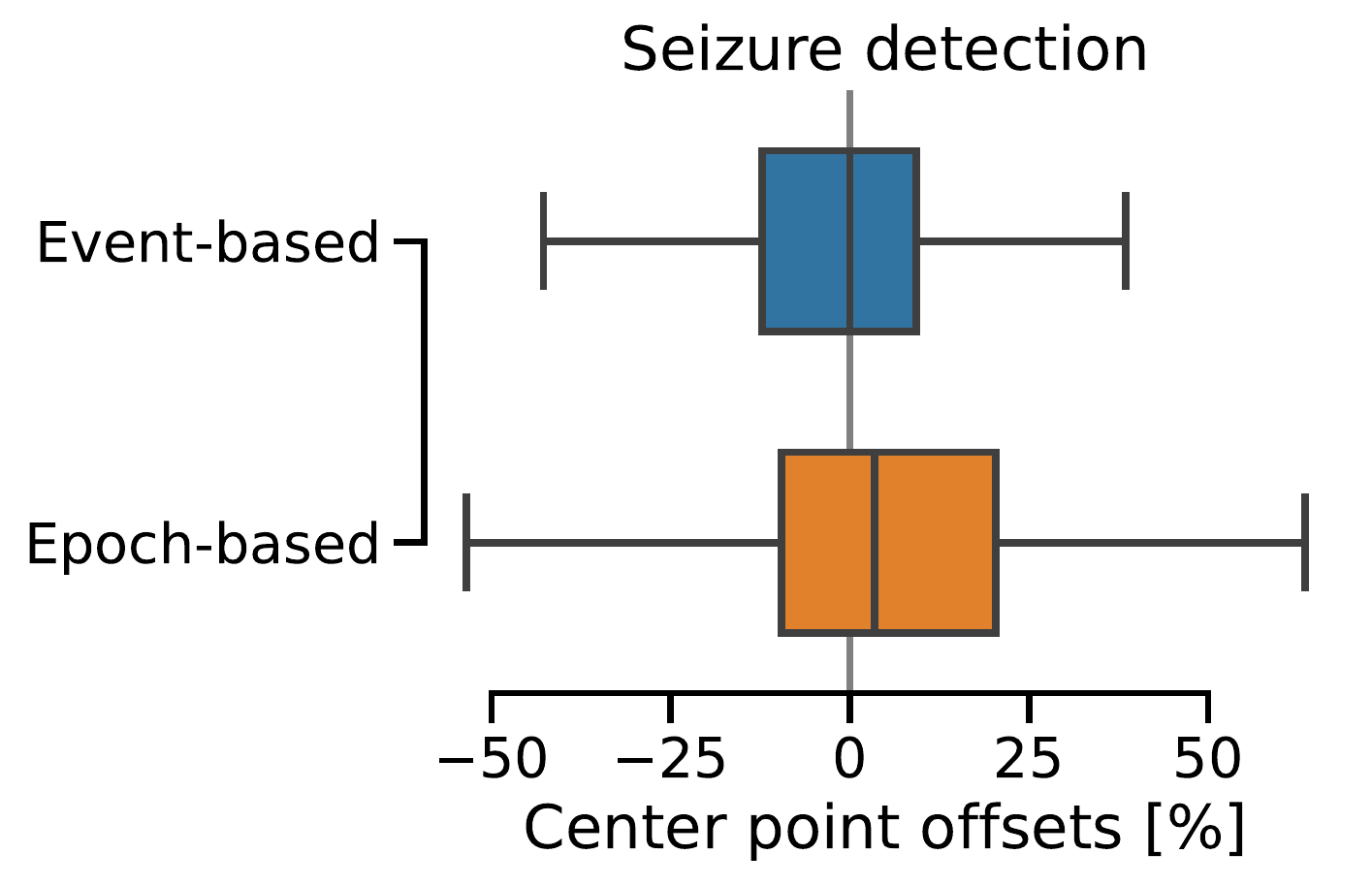}
		\caption{}
		\label{fig:tusz_center}
	\end{subfigure}
	\caption{Relative offsets between ground-truth and predicted centers (normalized by ground-truth duration). Positive values indicate that the predicted event center lies later in time than the ground truth. We consider all matched ground-truth and predicted events that show any overlap (green + orange class in Fig. \ref{fig:precrec})}
	\label{fig:center}
\end{figure*}

Regarding the center points and duration evaluation, both approaches are evaluated at the operating point corresponding to the maximum common recall level (0.4 for the artefacts, and 0.33 for the seizures). Relative offsets between ground-truth and predicted centers are shown in Figure \ref{fig:center} for both data sets. For artefact detection, our event-based approach shows less variability around ground-truth center points compared to the epoch-based one. For seizure detection, the event-based approach shows less variability, with the median predicted center being closer to ground-truth.

\begin{figure*}[t]
	\centering
	\begin{subfigure}[t]{0.4\textwidth}
		\centering
		\includegraphics[width=\textwidth]{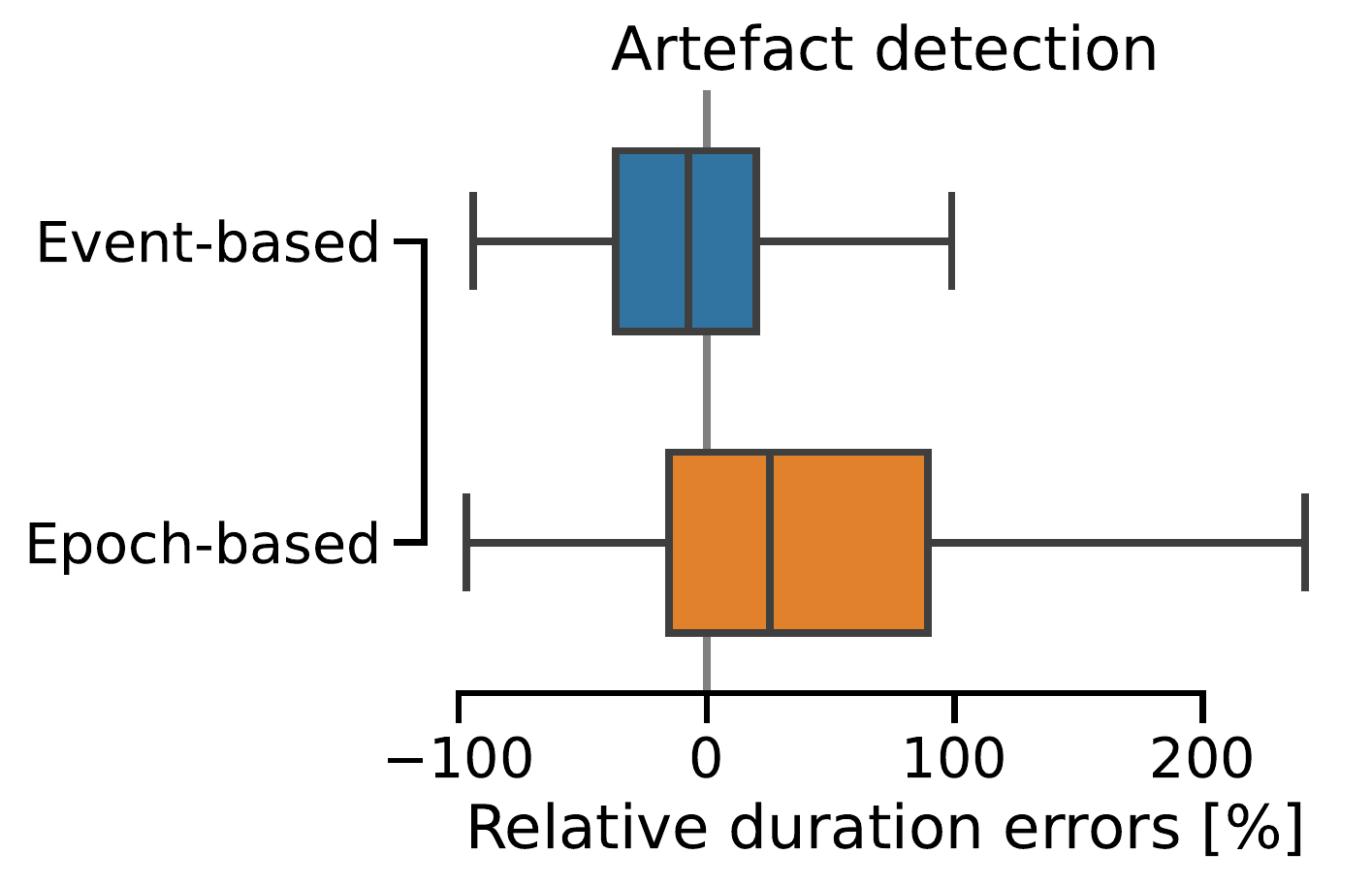}
		\caption{}
		\label{fig:tuar_dur}
	\end{subfigure}
	\hfill
	\begin{subfigure}[t]{0.4\textwidth}
		\centering
		\includegraphics[width=\textwidth]{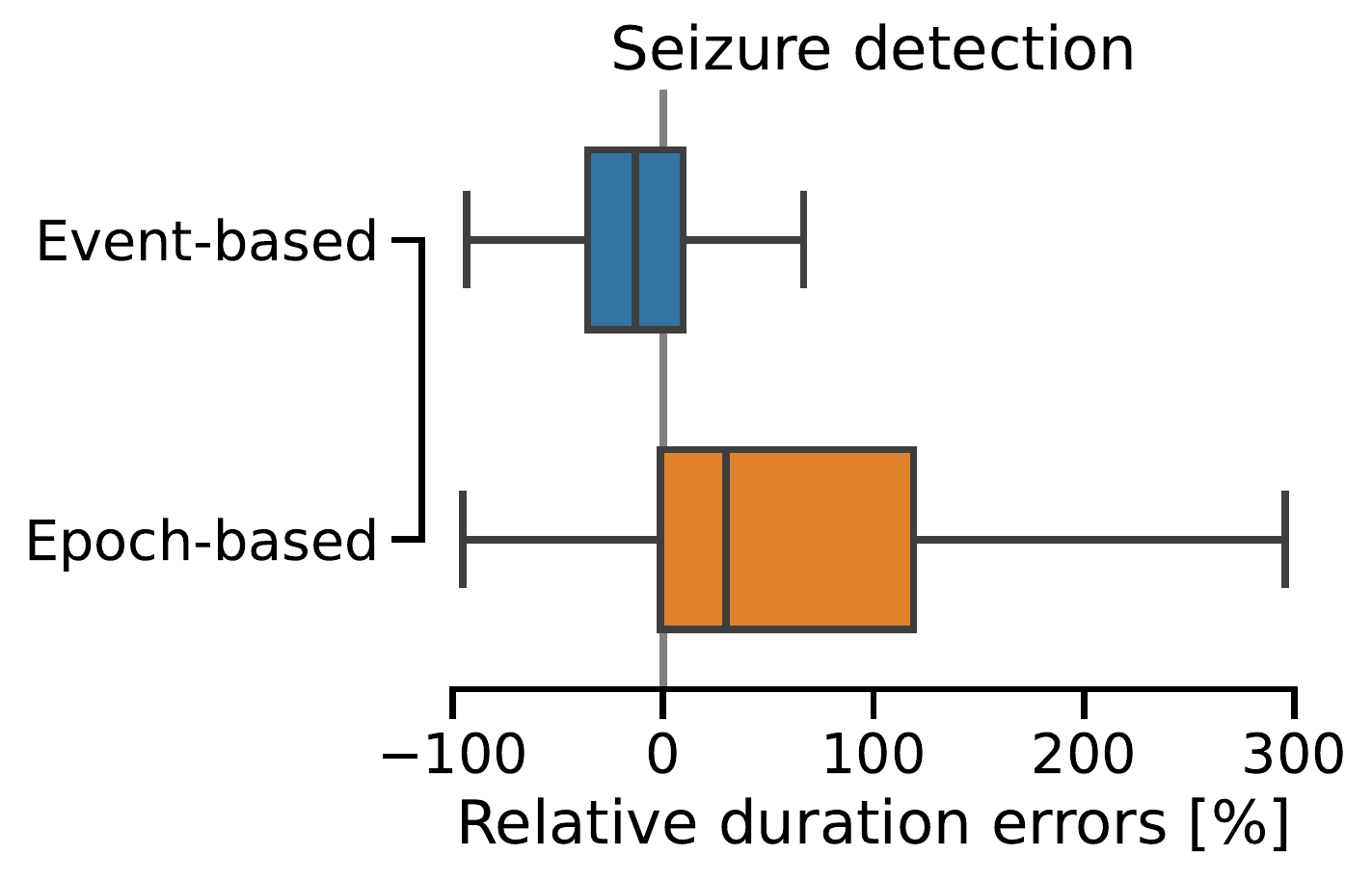}
		\caption{}
		\label{fig:tusz_dur}
	\end{subfigure}
	\caption{Relative errors between ground-truth and predicted durations (normalized by ground-truth duration). We consider all matched ground-truth and predicted events that show any overlap (green + orange class in Fig. \ref{fig:precrec})}
	\label{fig:dur}
\end{figure*}

Relative duration prediction errors are shown in Figure \ref{fig:dur} for both data sets. For artefact detection, the event-based median duration prediction is closer to the ground truth. Additionally, the event-based approach makes smaller errors in duration predictions (indicated by the lower interquartile range).

\section{Discussion}
\label{section_discussion}
\subsection{Discussion of experimental results}
Using the simulated event data set, we show the power of event-based modeling. The event-based approach clearly outperforms the epoch-based approaches. We want to emphasize that we do not claim superior performance in general, only strictly with these approaches. Our main goal is to show the ease with which a performant model can be designed with event-based modeling compared to epoch-based modeling.

On real-world data, the event-based approach outperforms the epoch-based benchmarks. Looking at the EEG artefact results, the event-based approach shows improved precision at all recall levels (Note that this is a very challenging data set). For the artefacts, additionally, our event-based approach shows improved duration prediction, based on the higher proportion of "True positives" relative to "<0.5IoU" in Figure \ref{fig:tuar_sampled} and in the prediction error evaluation of Figure \ref{fig:tuar_dur}. 

Regarding EEG seizures, the event-based approach generally outperforms the epoch-based benchmark. Only for 0.1 recall, the epoch-based benchmark manages to find slightly more events (as evidenced by the combination of the green and orange part of Figure \ref{fig:tusz_sampled}). At the same time, the event-based predictions are of slightly higher quality (as evidenced by the higher proportion of "True positives", i.e., higher proportion of >0.5 IoU overlap). Note that this epoch-based benchmark recently scored top place in a competition on this exact data set, and is highly tuned. Our event-based approach uses the base model of the benchmark and turned it into an event-based model by the addition of a center and duration "head".

Despite the potential benefit of performance due to an event-based approach, we want to emphasize that the largest benefit of our framework lies in removing the need for post-processing in epoch-based approaches. This post-processing was different for the three data sets, and will be different again for new data sets.

\subsection{Using an event-based framework}
We have proposed an event-based deep learning framework for time series, and have applied it to simulated ECG artefacts, and real-world artefact and seizure detection in EEG. Due to the end-to-end nature of our approach, and the neural network backbone learning its own data representation, our framework is more broadly applicable than these use cases. 

The major difference, and benefit, of event-based modeling compared to classical epoch-based approaches, is the lack of case-specific post-processing. Using event-based modeling, one can go directly from model output to events, whereas epoch-based approaches will always need some kind of tailored translation step before events can be listed. End-to-end event-based models drop the requirement for domain-specific post-processing rules. These can involve the expected event duration, duration ranges, how quickly events can follow each other, etc. While event-based models might also benefit from such post-processing, it is not a crucial step as opposed to existing approaches. Our framework can automatically learn most of those patterns from data, as is evident from our experiments. It should be noted that we do not claim that an event-based approach will always learn all patterns and will always outperform epoch-based approaches with well-designed post-processing. When extensive domain expertise is available, epoch-based approaches relying on features and post-processing rules inspired by this expertise can potentially outperform a generic event-based model, especially when limited training data is available.

One of the major benefits of working within an event-based framework, compared to segmenting signals into epochs, is the more intuitive nature of labels. The way our framework uses event labels matches closely to how human annotators would work with these labels (defining a start and stop time, which is equivalent to a center point and duration). This close match allows for easier feedback from human experts when developing a machine learning solution. This should be compared to the classical epoch-based approach, where labeled events need to be translated into a sequence of classification targets (with potential ambiguity at event borders), and back to events at inference time. This can add substantial friction to the development process, and hinder expert feedback.

A potential limitation or difficulty in using an event-based framework is the need for training data. Relevant features and event characteristics are learned jointly, relying on enough training examples to do so. A key aspect about these training examples is the diversity of durations. Our framework learns to directly predict durations, so it requires a broad range of example durations to learn from. The real-world artefact and seizure data sets both cover a wide range of durations, but are heavily skewed towards shorter events. The impact of duration distribution on performance is unknown at present. One can imagine, for example, that shorter durations are easier to predict if shorter events are more consistent in nature than longer events but this remains to be investigated.

\section{Conclusion}
\label{section_conclusion}
In this paper, we have proposed and showcased an event-based approach to a broad class of event detection problems in biomedical signal processing. The model can directly detect events of variable duration in long signal recordings. In contrast to existing epoch-based methods, we require no post-processing scheme to translate predictions into a set of events. Our model can be extended to other biomedical event-detection tasks and to other signal processing tasks where signal events are involved.

\printbibliography

\appendix
\setcounter{table}{0}
\renewcommand{\thetable}{A\arabic{table}}
\setcounter{figure}{0}
\renewcommand{\thefigure}{A\arabic{figure}}

\subsection{Simulated data generation} \label{app:generation}
Background signals are \SI{20}{\second} segments taken from the Computing in Cardiology 2017 Challenge signals \cite{clifford2017af}. For training data, the Challenge training set is used, for testing the Challenge validation set. The segments are drawn from the original data set with a \SI{5}{\second} stride.

For every background segment, target events are added in randomly with 20\% chance (to reflect a natural sparsity of target events common in biomedical signal processing tasks), taken from the Physionet MIT-BIH Noise Stress Test Database \cite{moody1984noise}. This process happens once for every background segment at the start of training, and the resulting signals and targets are used as finite data sets for experiments. In the cases where targets are added, either one or two segments (chosen randomly) are taken from the noise database of random duration, uniformly distributed between \SI{1}{\second} and \SI{6.7}{\second}. The target events are added to the background signal with a signal-to-noise ratio (treating the target artefact events as noise) uniformly random between -6 and 6. Target events for the training and test set are not taken from the same noise recording. To add more of a challenge to this simulation, additional artefact events are added to the background signal (in a similar fashion) with shorter duration, ranging between \SI{0.5}{\second} and \SI{1}{\second}.

\subsection{Network and training details} \label{app:details}
\subsubsection{Simulated events} \label{app:sim_details}
Simulated events are detected based on a single-channel input. The backbone's base "building block" is constructed using a 1D (temporal) convolution, a batch-normalization layer, and an ELU activation function. Every down- and upsample step is taken with a factor 4. The full backbone architecture is shown in Table \ref{tab:sim_backbone}. Two convolution layers with kernel size 7 are used in the event-based model on the output at stage 2' to produce the center and duration signals. A single convolution layer with kernel size 7 is used by the epoch-based models to produce their outputs. The networks are trained on input signal segments of \SI{20}{\second} (as explained above). The event-based loss function is described in the main text. The epoch-based loss function is binary cross-entropy.

\begin{table}[t]
	\centering
	\caption{Simulated events backbone. The different stages are connected in a U-Net-like way, stage 4 for example concatenates the features of stage 4 and upsampled features of stage 5. Every stage besides 0 contains a batch-normalization layer between the convolution and nonlinearity. Stages $()'$ have two convolution-normalization-nonlinearity blocks. The \textit{stride factor} indicates the total "downsample factor" for the given stage. This backbone produces a features signal at 1/16th the original sampling frequency.}
	\label{tab:sim_backbone}
	\begin{tabular}{c || c | c | c}
		Stage 	& \#Filters & Kernel size & Stride factor \\ \hline
		0 	& 32 	& 20 	& 1 \\ 
		1 	& 64 	& 20 	& 4 \\ 
		2 	& 64 	& 15 	& 16 \\ 
		3 	& 64 	& 15 	& 64 \\ 
		4 & 64 & 10 & 256 \\ 
		5' & 64 & 5 & 1024 \\ 
		4' & 64 & 10 & 256 \\ 
		3' & 64 & 15 & 64 \\ 
		2' & 64 & 15 & 16
	\end{tabular}
\end{table}

\subsubsection{EEG artefacts}
Artefact events are detected based on a single-channel input. The backbone's base "building block" is constructed using a separable 1D convolution, a batch-normalization layer, and an ELU activation function. Every down- or upsample step is taken with a factor 4. At the original input resolution (before the first downsampling step), the networks are using one such block, at every other stage using two such blocks (two at the "downward" path, and two at the "upward" path). The full backbone architecture is shown in Table \ref{tab:tuar_backbone}. Two "size 1" convolutions are used in the event-based model on the output at stage 2' to produce the center and duration signals. A single "size 1" convolution is used by the epoch-based model on the output at stage 2' to produce its output. Both networks are trained on input signal segments of \SI{200}{\second} and tested using a full EEG channel recording. The event-based loss function is described in the main text. The epoch-based loss function is binary cross-entropy with label smoothing, similar to the U-Net of \cite{neureka}.

For epoch-based post-processing, a median filter with a size corresponding to \SI{0.1}{\second} is used. The shortest event in the artefact data set lasts \SI{0.2}{\second}, so it is considered a reasonable filter size. The filter is used \textit{after} thresholding the model at a certain operating threshold.

\begin{table}[t]
	\centering
	\caption{Artefact detection backbone. The different stages are connected in a U-Net-like way, stage 5 for example concatenates the features of stage 5 and upsampled features of stage 6. Every stage besides 0 contains two separable convolutions with the given hyper-parameters. Stages $()'$ have a dropout layer between the two convolutions. The \textit{stride factor} indicates the total "downsample factor" for the given stage. This backbone produces a features signal at 1/16th the original sampling frequency.}
	\label{tab:tuar_backbone}
	\begin{tabular}{c || c | c | c}
		Stage 	& \#Filters & Kernel size & Stride factor \\ \hline
		0 	& 32 	& 20 	& 1 \\ 
		1 	& 64 	& 20 	& 4 \\ 
		2 	& 64 	& 15 	& 16 \\ 
		3 	& 64 	& 15 	& 64 \\ 
		4 & 64 & 10 & 256 \\ 
		5 & 64 & 5 & 1024 \\ 
		6 & 64 & 5 & 4096 \\ 
		5' & 64 & 5 & 1024 \\ 
		4' & 64 & 10 & 256 \\ 
		3' & 64 & 15 & 64 \\ 
		2' & 64 & 15 & 16
	\end{tabular}
\end{table}

\subsubsection{EEG Seizures}
The seizure detection backbone is constructed following the network of \cite{neureka}. This backbone is made up of a channel-independent encoder, and channel information is only merged at the deepest part of the backbone and the skip connections. Attention Gating \cite{schlemper2019attention} is used in the skip connections to merge channel information. The original network is only used until stage "4", at 1/256th of the original sampling frequency. After this stage, the event-based framework's center and duration heads are appended. This is done to limit computational and memory footprint, since there is little performance impact compared to applying the heads at the original sampling frequency (as is done in the U-Nets of \cite{neureka}). The full backbone architecture is shown in Table \ref{tab:tusz_backbone}. The epoch-based seizure detector is the original model of \cite{neureka}, with an output at the same resolution as the original input. Similar to artefact detection, the seizure detection networks are trained on input segments \SI{200}{\second}, and tested on full recordings. 

For epoch-based post-processing, a median filter with a size corresponding to \SI{2}{\second} is used. Similar to the artefact detection setting, this filter is applied after thresholding the model's output.

\begin{table}[t]
	\centering
	\caption{Seizure detection backbone. Stage 4 and 4' are connected in a U-Net-like way, concatenating the features of stage 4 and upsampled features of stage 5''. Every stage contains convolution layers with the given hyper-parameters. The  $()'$ stages have two convolution-normalization-nonlinearity blocks with the given parameters. Stage 4' combines the features of stage 4 using Attention Pooling \cite{schlemper2019attention} before processing. Stage 5' has a dropout layer after its two convolution blocks and performs max-pooling over the different EEG channels before processing. All encoder stages are channel-independent (using the same convolution filter on every channel), the decoder stages merge channel-level information. The \textit{stride factor} indicates the total "downsample factor" for the given stage. This backbone produces a features signal at 1/256th the original sampling frequency.}
	\label{tab:tusz_backbone}
	\begin{tabular}{c || c | c | c}
		Stage 	& \#Filters & Kernel size & Stride factor \\ \hline
		0 	& 16 	& 15 	& 1 \\ 
		1 	& 32 	& 15 	& 4 \\ 
		2 	& 64 	& 15 	& 16 \\ 
		3 	& 64 	& 7 	& 64 \\ 
		4 & 128 & 3 & 256 \\ 
		5 & 128 & 3 & 1024 \\ 
		5' & 64 & 3 & 1024 \\ 
		4' & 64 & 5 & 256
	\end{tabular}
\end{table}

\subsection{Precision-recall curves for simulated events} \label{app:precision_recall}
\begin{figure}
	\centering		\includegraphics[width=\linewidth]{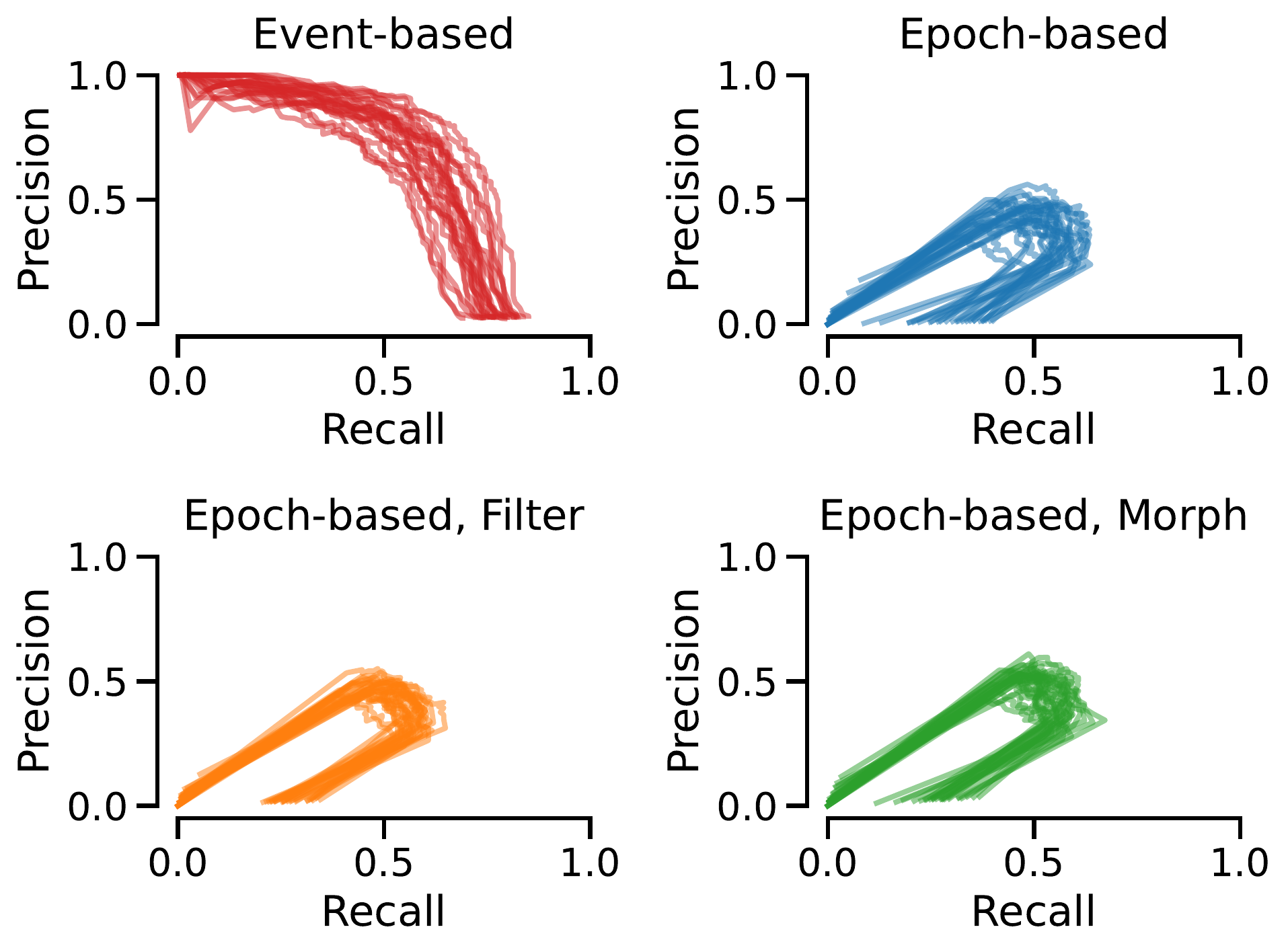}
	\caption{Precision-recall curves for the detection of simulated events, evaluated with a 0.75 IoU threshold criterion. Every line is a separate run (data generation + training)}
	\label{fig:precision_recall}
\end{figure}

Full precision-recall curves for the different approaches used for simulated events (evaluated with a 0.75 IoU threshold criterion) can be found in Figure \ref{fig:precision_recall}. These curves are produced by sweeping over different operating points, i.e., sweeping the range [0, 1] on the models' output confidence. The event-based approach produces a traditional, intuitive precision-recall curve with high precision and low recall at one end of the curve, and low precision combined with high recall at the other end. The epoch-based approaches do not show this behavior and seem to have a single Pareto-optimal operating point (also observed for the real-world EEG artefacts in Figure \ref{fig:tuar_sampled}).

\subsection{Stable focal loss} \label{app:stable}
The softplus function is a part of focal loss. Instabilities are expected when computing the gradient of the focal loss because of this softplus function. The softplus function can be found in the "cross-entropy factor" of the focal loss formulation (taking the logarithm of a sigmoid activation): $$\log \sigma (l(x)) = \log e^l - \log (e^l + 1) = l - \text{softplus}(l)$$ and $$\log (1 - \sigma(l(x))) = \log (1 - \frac{e^l}{e^l + 1})$$ $$= \log 1 - \log(e^l + 1) = - \text{softplus}(l)$$ 

The softplus function $\log(e^{(\cdot)} + 1)$ can be rewritten in a numerically more stable form that avoids a possibly exploding gradient as follows: $$\text{softplus}_{stable}(x) = \log(e^{-|x|} + 1) + \max(x, 0)$$

\end{document}